\documentclass[12pt]{article}
\usepackage[latin1]{inputenc}
\usepackage[dvips]{graphicx}
\usepackage{graphicx}
\newcommand{\text}{\rm}

\newcommand{\ug}{ \; = \; }

\newcommand{\infi}{\infty}

\newcommand{\bb}{\begin{equation}}
\newcommand{\ee}{\end{equation}}
\newcommand{\bega}{\begin{eqnarray}}
\newcommand{\ega}{\end{eqnarray}}
\newcommand{\begae}{\begin{eqnarray*}}
\newcommand{\egae}{\end{eqnarray*}}

\newcommand{\h}{\hspace*{4ex}}
\newcommand{\dis}{\displaystyle}

\newcommand{\cent}{\centerline}
\newcommand{\vs}{\vspace*}

\begin{document}

\baselineskip 0.8cm

\begin{center}

{\large {\bf Analytic description of Airy-type beams when truncated by finite apertures} $^{\: (\dag)}$ }
\footnotetext{$^{\: (\dag)}$ E-mail addresses for contacts: mzamboni@dmo.fee.unicamp.br}

\end{center}

\vs{5mm}

\cent{Michel Zamboni-Rached}

\vs{0.2 cm}

\centerline{{\em DMO--FEEC, Universidade Estadual de Campinas, Campinas, SP, Brasil.}}

\vs{0.3 cm}

\cent{K. Z. N\'obrega}

\vs{0.2 cm}

\cent{{\em Departamento de Eletro-Eletr\^onica, Instituto Federal do Maranh\~ao, São Luis, Ma, Brasil}}

\vs{0.2 cm}

\centerline{\rm and}

\vs{0.3 cm}

\cent{C. A. Dartora}

\vs{0.3 cm}

\cent{{\em Departamento de Engenharia El\'etrica, Universidade Federal do Paran\'a, Curitiba, Pr, Brasil}}

\vs{0.5 cm}


{\bf Abstract  \ --} \ In this paper, we have developed an analytic method for describing Airy-Type beams
truncated by finite apertures. This new approach is based on suitable superposition of exponentially decaying
Airy beams. Regarding both theoretical and numerical aspects, the results here shown are interesting because
they have been quickly evaluated through a simple analytic solution, whose characteristics of propagation has
agreed with those already published in literature through the use of numerical methods. To demonstrate the
method's potentiality, three different truncated beams have been analyzed: ideal Airy, Airy-Gauss and
Airy-Exponential.


\section{Introduction}

\h In the last five years, a new nondiffracting wave has called attention: the Airy Beam[1-4]. Solution of the
wave equation under paraxial approximation, the Airy Beam has slightly different properties. Contrary to Bessel
and Mathieu beams that keep their transverse shapes only, the Airy beam maintains that property but also it is
observed that its main spot propagates according to a parabolic trajectory, which gives the idea of a bent
propagation or a self-accelerated beam.

\h Like every ideal nondiffracting wave, the ideal Airy Beam can propagate over infinite distance resisting the
diffraction effects, but it presents an infinite power flux through any plane orthogonal to the propagation
direction.

\h To overcome this problem, Siviloglou and Christodoulides \cite{cris1} obtained a finite energy Airy beam
solution given by an ideal Airy beam modulated by an exponentially decaying function at $z=0$.

\h Another possibility is to perform a spatial truncation on the ideal Airy beam (finite aperture generation).
Actually, the spatial truncation is the most effective and realistic option, since every beam must be generated
by finite apertures.

\h Until now, to the best of our knowledge, no one has ever studied propagation characteristics of truncated
Airy-Type beams using any kind of analytic method once that all results reported in literature were done
numerically \cite{car} or experimentally \cite{cris2}. In this paper we will present the first effort to this
direction, describing the propagation of truncated Airy-Type beams in a homogeneous medium through an analytic
approach.

\h The method here described is based on suitable superposition of exponentially decaying Airy beams. To support
the method three different truncated beams have been analyzed: ideal Airy, Airy-Gauss and Airy-Exponential.

\h The results here presented are of interest in any possible application, theoretical or practical, that makes
use of Airy-Type beams.

\section{Analytic Description for Truncated Airy-Type Beams}

\h In \cite{cris1}, Siviloglou et al. considered an initial field profile, at $\zeta=0$, given by

 \bb \psi(s,\zeta=0) \ug Ai(s)\exp(as) \label{iab} \ee

\

as an initial condition to the electric field envelope equation in the paraxial regime and (1+1)D,
$i\partial_{\zeta}\psi + 1/2\partial^2_{s}\psi =0$, obtaining the following finite energy Airy beam

 \bb \psi(s,\zeta) \ug Ai(s - (\zeta/2)^2 + ia\zeta)\exp(as - (a\zeta^2/2) -i\zeta^3/12 + ia^2\zeta/2 + is\zeta/2   )\,\, , \label{aeb} \ee

\

where $s = x/x_0$ and $\zeta = z/kx_0^2$ are the dimensionless transverse and longitudinal coordinates, with
$x_0$ being the spatial spotlight and $k=2\pi n/\lambda_0$ the wavenumber of the optical wave. We shall named
the solution (\ref{aeb}) as Airy-Exponential beam.

\h Before expose our method, it is important to notice that our goal is to describe, \emph{analytically},
Airy-Type beams truncated by finite apertures at $\zeta=0$, i.e., we wish a solution describing (approximately)
the evolution of fields of the type $\Psi(s,\zeta) = Ai(s)\, m(s)[\theta(s+S) - \theta(s-S)]$, where $m(s)$ is a
function that \emph{modulates the Airy function} within the finite aperture, $-S\leq s \leq S$, which is
represented by the difference between the Heaviside functions $\theta(s+S)$ and $\theta(s-S)$. Here, $S = X/x_0$
is the dimensionless width of the aperture.


\h Now we start to develop our method by considering the initial field profile, at $\zeta=0$, as the following
superposition

\bb \Psi(s,\zeta=0) \ug \sum_{n=-\infi}^{\infi} B_n\,Ai(s)\exp(a_n s)\,\, , \label{psi0} \ee

\

where $B_n$ and

$a_n$ are complex constants.

\h Of course the resulting field, $\Psi(s,\zeta)$, emanated from the plane $\zeta=0$, will be given by the
superposition of Airy-Exponential beams:

\bb \Psi(s,\zeta) \ug \sum_{n=-\infi}^{\infi} B_n\,Ai(s - (\zeta/2)^2 + ia_n\zeta)\exp(a_n s - (a_n\zeta^2/2)
-i\zeta^3/12 + i\,a_n^2\zeta/2 + is\zeta/2 ) \,\, . \label{PSI}\ee

\h Let us return to the initial field (\ref{psi0}) and make the choice

\bb a_n = a_R + i 2\pi\,n/L \,\, , \label{an}\ee

\

where $a_R$ and $L$ are \emph{positive} constants. It is important to notice that $a_R$ is the same for all
terms in the summatory. Therefore, the initial field is rewritten as:

\bb \Psi(s,\zeta=0) \ug  Ai(s)\exp(a_R\,s)\sum_{n=-\infi}^{\infi} B_n \exp(i\frac{2\,\pi}{L}n\,s) \label{psi02}
\ee

\

and we define a function $I$ as the whole expression of summatory in eq.(\ref{psi02}),

\bb I(s) \ug \sum_{n=-\infi}^{\infi} B_n \exp(i\frac{2\,\pi}{L}n\,s) \,\,\, , \label{I} \ee

\

that clearly is a Fourier's Series with $L$ as the period.

\h Now, we chose $I(s)$, within $-L/2 \leq s \leq L/2$, as given by:

\bb
 I(s) \ug \left\{\begin{array}{clr}
 \exp(-a_R\,s)m(s);\;\; & {\rm for}\;\;\; -S \leq s \leq S  \\
\\
 \;\;\;\; 0;\;\;  & {\rm for}\;\;\; -L/2 \leq s \leq -S \,\, {\rm and}\,\, S \leq s \leq L/2

\end{array} \right. \label{I0}
 \ee

 \

 that results on the following $B_n$ in eqs.(\ref{I},\ref{psi02}):

 \bb B_n \ug \frac{1}{L} \int_{-S}^{S} \exp(-a_R\,s)m(s)\exp(-i\frac{2\,\pi}{L}n\,s) ds    \label{bn} \ee
 where, as we are going to see, $m(s)$ is a function that will modulate the Airy function within the finite aperture.

\h The reason for choosing $I(s)$, eq.(\ref{I}), as given by (\ref{I0}) will become clear in the next step.

\h By using our choices (\ref{an}) and (\ref{bn}) into eq.(\ref{psi02}) we get the following expression to the
initial field:

\bb \begin{array}{l} \Psi(s,\zeta=0) =  Ai(s)\exp(a_R\,s)\dis{\sum_{n=-\infi}^{\infi}} B_n
\exp(i\frac{2\,\pi}{L}n\,s) \\

\\

 = \left\{\begin{array}{clr}
 Ai(s)\,m(s)  & {\rm for}\;\; |s| \leq S  \\
\\
  0  & {\rm for}\;\; S < |s| \leq L/2 \\
 \\
 Ai(s)\,\exp(a_R\,s)\,I(s)\approx 0 & {\rm for}\;\; |s| \geq L/2

\end{array} \right. \end{array}\label{psi03}
 \ee

\

where $I(s)$ is the same of eq.(\ref{I}) and so it repeats its values in periodic space intervals. Since
$L/2>S$, \emph{for appropriated choices} of $L$ and $a_R$ we have that $Ai(s) \exp(a_R\,s)I(s)\approx 0$ for
$|s| \geq L/2$ due to the behavior of the functions $Ai(s)$ and $\exp(a_R s)$ for positive and negative values
of $s$, respectively.

\h In this way, we have shown that the initial field given by (\ref{psi0}), with $a_n$ and $B_n$ given by
(\ref{an}) and (\ref{bn}), can represent at $z=0$ an Airy-Type beam truncated by a finite aperture.

\h Finally, once the truncated Airy-Type beam at $z=0$ is described by eq.(\ref{psi0}), the resulting beam
emanated from the finite aperture will be given by the solution (\ref{PSI}).

\section{Applying the method to three truncated Airy-Type beams}

\h In this Section we shall apply our method to three situations involving Airy-Type beams truncated by finite
apertures: the ideal Airy beam, the Airy-Exponential beam and the Airy-Gauss beam. Obviously, we will use a
finite number $2N+1$ in (\ref{psi0}) and (\ref{PSI}), with $-N \leq n \leq N$.

\h It is important to notice that the choice of the values of $L$ and $a_R$ in (\ref{an}) is not unique,
actually there are many alternative sets of those values that yield excellent results.

\h In all cases we shall assume a wavelength of 500 nm.

\subsection{Analytic description of the truncated Ideal Airy beam}

\h Let us consider an Ideal Airy beam truncated, at $\zeta=0$ (i.e., at $z=0$), by a linear aperture of width
$2X$; i.e., $\Psi(s,\zeta=0)=Ai(s)\,(\theta(s+S)-\theta(s-S))$, where we chose (the spot size) $x_0=100\,\mu$m
and $X=1.65\,$mm. We remember that $s=x/x_0$, $\zeta=z/kx_0^2$ and $S=X/x_0$.

\h At $\zeta=0$ this field is described by eq.(\ref{psi0}), with $a_n$ and $B_n$ given by eqs.(\ref{an}) and
(\ref{bn}) respectively and with $m(s)=1$. In this example, an excellent result can be obtained by the choice $L
=50$, \ $a_R = 0.1$  and $N=60$.


\h Figure 1 shows the intensity of the field given by eq.(\ref{psi0}) and we can see that it represents, at
$\zeta=0$, the truncated Ideal Airy beam with high fidelity

\h The field emanated by the aperture is given by solution (\ref{PSI}), and its intensity is shown in Fig.2a.
This result corresponds to an Ideal Airy beam truncated by a finite aperture. Figure 2b shows the orthogonal
projection of this case.

\begin{figure}[!h]
\begin{center}
 \scalebox{2}{\includegraphics{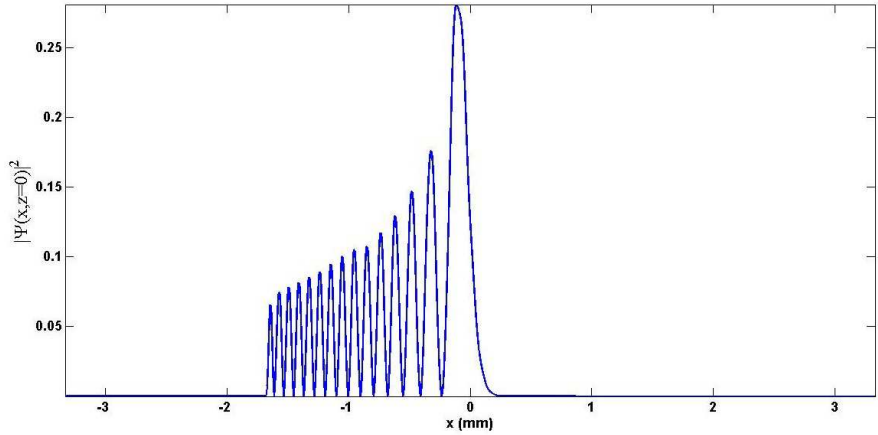}}
\end{center}
\caption{Field given by eq.(\ref{psi0}), representing an Ideal Airy beam, with $x_0=100\,\mu$m and truncated, at
$\zeta=0$, by a linear aperture of width $2X=3.3\,$mm. The quantities $a_n$ and coefficients $B_n$ are given by
eqs.(\ref{an},\ref{bn}), with $m(s)=1$. We use $a_R=0.1$, \ $L = 50$, and $N=60$.} \label{fig1}
\end{figure}

\begin{figure}[!h]
\begin{center}
 \scalebox{2.2}{\includegraphics{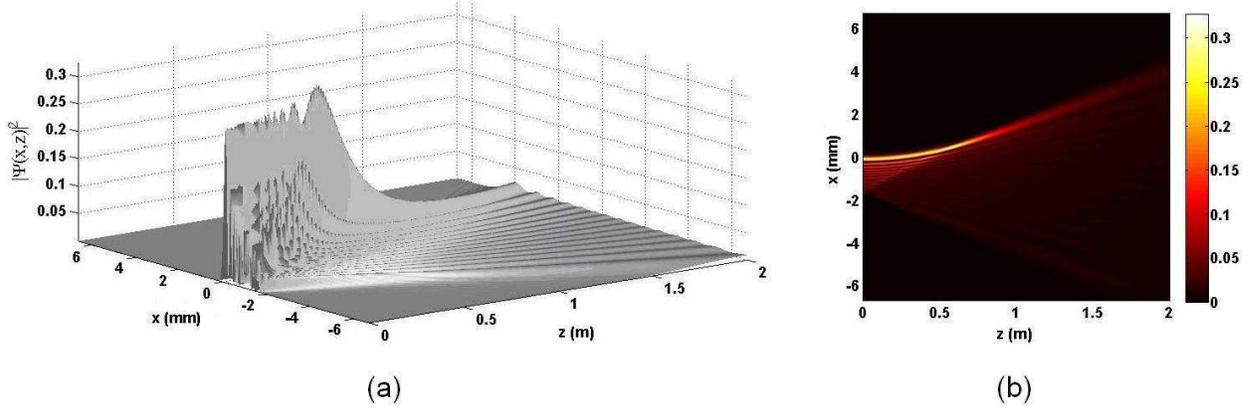}}
\end{center}
\caption{(a)Intensity of an Ideal Airy beam truncated by a finite aperture, as given by solution (\ref{PSI}).(b)
The same result depicted in orthogonal projection} \label{fig2}
\end{figure}

\h In spite of the excellent results, we could get more accurate solutions by increasing the number of terms in
the series (\ref{PSI}), which expresses the resulting field, while keeping the same values for $L$ and $a_R$.

\

\subsection{Analytic description of the truncated Airy-Exponential beam}

\h Let us consider an Airy-Exponential beam truncated, at $\zeta=0$, by the same aperture of width $2X$; i.e.,
$\Psi(s,\zeta=0)=Ai(s)\,\exp(q s)\,(u(s+S)-u(s-S))$. Here we also chose $x_0=100\,\mu$m and $X=1.65\,$mm, with
the value of $q$ set as $q = 0.05$.

\h At $\zeta=0$ the field is approximately described by eq.(\ref{psi0}), with $a_n$ and $B_n$ given by
eqs.(\ref{an}) and (\ref{bn}) respectively and with $m(s)=\exp(q s)$. Here we obtain quite good result by
choosing, again, $L =50$, \ $a_R = 0.1$ and $N=60$.


\h Figure 3 shows the intensity of the field given by eq.(\ref{psi0}) and we can see that it represents very
well, at $z=0$, the truncated Airy-Exponential beam.

\h The resulting beam emanated by the aperture is given by solution (\ref{PSI}), and its intensity is shown in
Fig.4a. Figure 4b shows the orthogonal projection of this case.

\begin{figure}[!h]
\begin{center}
 \scalebox{2}{\includegraphics{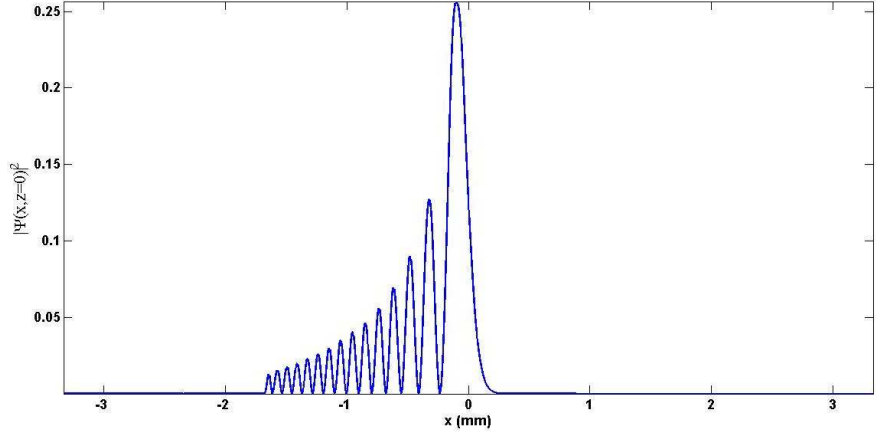}}
\end{center}
\caption{Field given by eq.(\ref{psi0}), representing, at $z=0$, an Airy-Exponential beam, with $q=0.05$ and
$x_0=100\,\mu$m, truncated by a linear aperture of width $2X=3.3\,$mm. The quantities $a_n$ and coefficients
$B_n$ are given by eqs.(\ref{an},\ref{bn}) with $m(s)=\exp(q s)$. We use $a_R=0.1$, \ $L = 50$, and $N=60$.}
\label{fig3}
\end{figure}

\begin{figure}[!h]
\begin{center}
 \scalebox{2.2}{\includegraphics{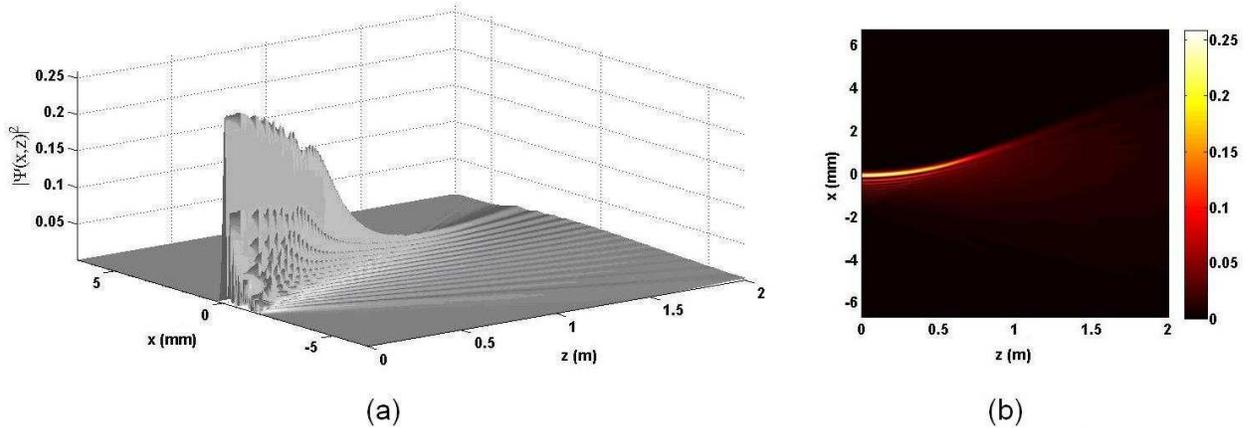}}
\end{center}
\caption{(a)Intensity of an Airy-Exponential beam truncated by a finite aperture, as given by solution
(\ref{PSI}).(b) The same result depicted in orthogonal projection} \label{fig4}
\end{figure}

\newpage

\subsection{Analytic description of the truncated Airy-Gauss beam}

\h Now, we wish to consider an Airy-Gauss beam truncated by a linear aperture of width $2X$; i.e.,
$\Psi(s,\zeta=0)=Ai(s)\,\exp(-q s^2)\,(\theta(s+S)-\theta(s-S))$, with the spot size $x_0=100\,\mu$m,
$X=1.65\,$mm and $q=5x_0^2/X^2=0.018$. This aperture possesses a size large enough to accommodate almost the
entire power-flux of the Airy-Gauss beam.

\h At $\zeta=0$ this field is described by eq.(\ref{psi0}), with $a_n$ and $B_n$ given by eqs.(\ref{an}) and
(\ref{bn}) respectively and with $m(s)=\exp(-q s^2)$. In this case, an excellent result can be obtained by the
choice $L =50$, \ $a_R = 0.1$ and $N=60$.

\h Figure 5 shows the intensity of the field given by eq.(\ref{psi0}) and we can see that it represents, at
$z=0$, the truncated Airy-Gauss beam with high fidelity


\begin{figure}[!h]
\begin{center}
 \scalebox{2}{\includegraphics{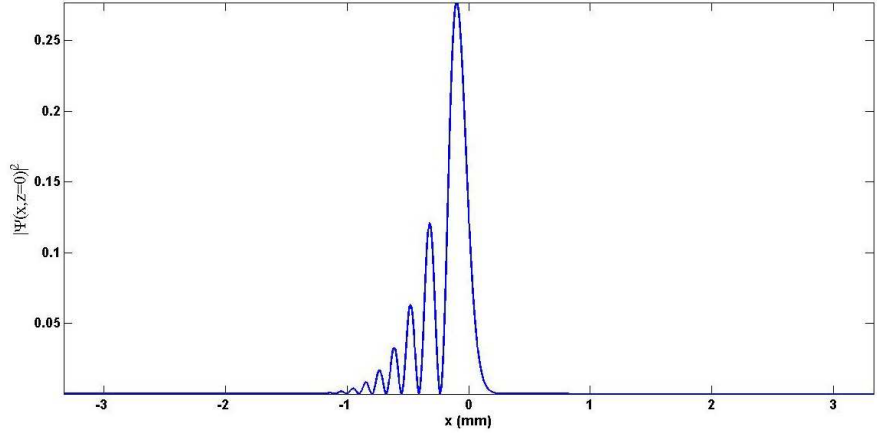}}
\end{center}
\caption{Field at $z=0$ given by eq.(\ref{psi0}), representing an Airy-Gauss beam, with $q=0.018$ and
$x_0=100\,\mu$m, truncated by a linear aperture of width $2X=3.3\,$mm. The quantities $a_n$ and coefficients
$B_n$ are given by eqs.(\ref{an},\ref{bn}) with $m(s)=\exp(-q s^2)$. We use $a_R=0.1$, \ $L = 50$, and $N=60$.}
\label{fig5}
\end{figure}

\begin{figure}[!h]
\begin{center}
 \scalebox{2.2}{\includegraphics{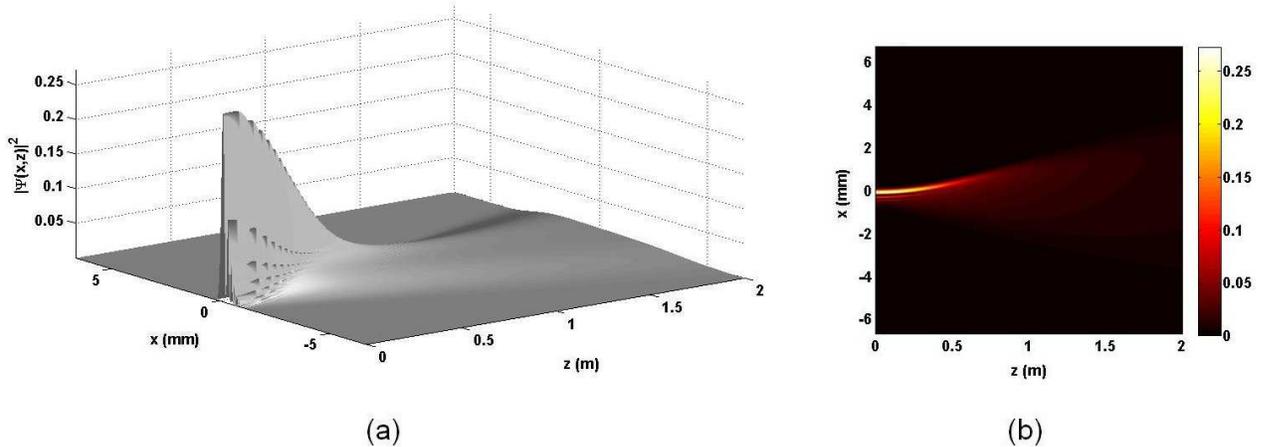}}
\end{center}
\caption{(a)Intensity of an Airy-Gauss beam truncated by a finite aperture, as given by solution (\ref{PSI}).(b)
The same result depicted in orthogonal projection} \label{fig6}
\end{figure}

\h The beam emanated by the aperture is given by solution (\ref{PSI}), and its intensity is shown in Fig.6a.
This result corresponds to an Airy-Gauss beam truncated by a finite aperture. Figure 2b shows the orthogonal
projection of this case.


\section{Acknowledgments}

\nopagebreak The authors are grateful to Erasmo Recami, Hugo E. Hern\'andez Figueroa, Jane M. Madureira Rached
and Suzy Zamboni Rached for many stimulating contacts and discussions. The authors acknowledge partial support
from FAPESP (under grant 11/51200-4); from CNPq (under grants 307962/2010-5 and 301079/2011-0).

\

\section{Conclusion}

\h In this paper we developed an analytic method capable to describe Airy-Type beams truncated by finite
apertures.

\h This new approach is based on suitable superposition of Airy-Exponential beams and presents a simple analytic
solution. The results agree with those already published in literature through the use of numerical methods. We
demonstrated the method's potentiality applying it to three different truncated Airy type beams: the Ideal Airy,
the Airy-Gauss and the Airy-Exponential, but many others Airy-Type beams can easily be described with this
analytic approach.

\end{document}